\documentstyle[aps,prl,epsf,multicol]{revtex}
\begin{document}
\title{\bf Statistics of Multiple Sign Changes in a Discrete Non-Markovian Sequence}

\author{Satya N. Majumdar}

\address{
Laboratoire de Physique Quantique (UMR C5626 du CNRS),
Universit\'e Paul Sabatier, 31062 Toulouse Cedex, France. \\}

\date{\today}

\maketitle

\begin{abstract} 
We study analytically the statistics of multiple sign changes
in a discrete non-Markovian sequence $\psi_i=\phi_i+\phi_{i-1}$ ($i=1,2,\dots,n$)
where $\phi_i$'s are independent and identically distributed random variables
each drawn from a symmetric and continuous distribution $\rho(\phi)$. We show
that the probability $P_m(n)$ of $m$ sign changes upto $n$ steps is
universal, i.e., independent of the distribution $\rho(\phi)$. The mean
and variance of the number of sign changes are computed exactly for all $n>0$.
We show that the generating function ${\tilde P}(p,n)=\sum_{m=0}^{\infty}P_m(n)p^m\sim 
\exp[-\theta_d(p)n]$ for large $n$ where the `discrete' partial survival exponent
$\theta_d(p)$ is given by a nontrivial formula, 
$\theta_d(p)=\log[{{\sin}^{-1}(\sqrt{1-p^2})}/{\sqrt{1-p^2}}]$ for $0\le p\le 1$.
We also show that in the natural scaling limit $m\to \infty$, $n\to \infty$ but keeping
$x=m/n$ fixed, $P_m(n)\sim \exp[-n \Phi(x)]$ where the large deviation function
$\Phi(x)$ is computed. The implications of these results for Ising spin glasses
are discussed.

\noindent

\end{abstract}

\begin{multicols}{2}

The probability $P_0(T)$ that a stochastic process $\psi(T)$ does not cross zero upto time $T$
is a quantity of long standing interest to both physicists and mathematicians\cite{BL,Slepian}
and has resurfaced recently with a new name `persistence' in the context of nonequilibrium 
systems\cite{Review}. A lot of recent efforts have been devoted to compute $P_0(T)$ for
stationary Gaussian processes. Such a Gaussian stationary process (GSP) is completely specified 
by its two
point correlation function $C(T)=\langle \psi(0)\psi(T)\rangle$. For a wide class
of correlation functions, it is known that $P_0(T)\sim \exp(-\theta T)$ for large $T$
where the persistence exponent $\theta$ is usually nontrivial, depends on the full
function $C(T)$ and is calculable exactly only in very few cases\cite{Review}. A
natural generalization of $P_0(T)$ is $P_m(T)$, the probability of $m$ zero crossings upto time $T$.
The mean and the variance of the number of zero crossings upto time $T$ of a GSP have been studied
before\cite{BL}. For a smooth GSP where $C(T)=1-aT^2+\ldots$ for small $T$ with $a>0$, the mean 
is given by Rice's formula\cite{Rice}, $\langle m\rangle/T=\sqrt{-C''(0)}/\pi$ and the variance
by a more complicated formula due to Bendat\cite{Bendat}. Recently it was shown\cite{MB} that for a 
smooth GSP the generating function 
${\tilde P}(p,T)=\sum_{m=0}^{\infty}P_m(T)p^m\sim \exp[-\theta(p)T]$
for large $T$ where the `partial survival' exponent $\theta(p)$ varies smoothly from $\theta(0)=\theta$
to $\theta(1)=0$ as $p$ varies continuously from $0$ to $1$. In Ref.\cite{MB}, the exponent $\theta(p)$
was computed exactly for the one dimensional Ginzburg Landau model of deterministic coarsening
and also approximately within the independent interval approximation for other smooth
processes such as the diffusion equation. The only process for which a closed form expression of 
$\theta(p)$ exists, so far, is the random acceleration problem\cite{Burkhardt,SGL} which corresponds
to a GSP\cite{Review} with correlator
$C(T)=[3\exp(-T/2)-\exp(-3T/2)]/2$. For this problem, the exponent $\theta(p)$ for $0\le p\le 1$
is given by the formula,
$\theta(p)={1\over {4}}\left[1-{6\over {\pi}}{\sin}^{-1}\left(p/2\right)\right]$.

In this Rapid Communication, we study the statistics of multiple crossings or sign changes
in a discrete sequence $\psi_1$, $\psi_2$, $\ldots$, $\psi_n$ as opposed to a continuous
process $\psi(T)$ discussed in the previous paragraph. This study is motivated by the 
recent works on the persistence of a discrete sequence\cite{MBE,MD,EB}. 
The principal motivation for
studying the persistence of a discrete sequence is twofold. First, in various experiments
and numerical simulations to measure the persistence $P_0(T)$ of a continuous stochastic process 
$\psi(T)$,
one usually samples the continuous process only at discrete time points separated by a fixed window size 
$\Delta T$ and checks whether
the process has retained the same sign at all these discrete times. Some information gets lost 
due to 
this discretization since the continuous process $\psi(T)$ may have crossed and recrossed zero in 
between
two successive discrete points. Thus the `discrete-time' persistence $P_0(n)$ [i.e., the probability
that the sequence $\psi(0)$, $\psi(\Delta T)$, $\psi(2\Delta T)$, $\ldots$, $\psi(n\Delta T=T)$
have the same sign]
is usually greater than the continuous time persistence $P_0(T)$. In Ref. \cite{MBE}, it was shown
that $P_0(n)\sim \exp[-\theta_d n]$ where the exponent $\theta_d$ depends continuously on the
window size $\Delta T$. The second motivation for studying the persistence of a sequence 
follows from the observation\cite{MD} that many processes in nature such as weather records are 
stationary
under translations in time only by an integer multiple of a basic period (which can be chosen to be 
unity without loss of generality). It was shown in Ref. \cite{MD}
that for a wide class of such processes, the continuous time persistence $P_0(T)$ is the same
as the persistence $P_0(n)$ of the corresponding discrete sequence obtained from the measurement of the 
process only at integer times. A natural generalization of $P_0(n)$ is clearly $P_m(n)$, the probability
that there are $m$ sign changes along a sequence of size $n$. 

The exact calculation of $P_m(n)$ for an arbitrary stationary sequence seems difficult. It is therefore
important to find exactly solvable cases. In 
this paper we present exact results for $P_m(n)$ for a specific sequence which was introduced 
in Ref. \cite{MD},
\begin{equation}
\psi_i = \phi_i +  \phi_{i-1},\,\,\,\,\, i=$1$,$2$,$\ldots$,$n$
\label{psi}
\end{equation}   
where $\phi(i)$'s are independent and identically distributed (i.i.d) random variables, not 
necessarily Gaussian,
each drawn from the same symmetric continuous distribution $\rho(\phi)$. The variables
$\psi_i$'s have only nearest neighbour correlations. The sequence in Eq. 
(\ref{psi}) is stationary but non-Markovian since $\psi_i$ depends
not just only on $\psi_{i-1}$ but on all the preceding members of the sequence\cite{MD}. This 
sequence appears as a limiting case
of the diffusion equation on a hierarchical lattice\cite{MD}. It also appears in the one dimensional
Ising spin glass problem where $\psi_i$ represents the energy cost to flip
the $i$-th spin\cite{DG}. In Ref. 
\cite{MD}, the persistence $P_0(n)$ for this sequence was computed exactly for all $n$ 
and remarkably $P_0(n)$ was found to be {\it universal}, i.e., independent of the distribution 
$\rho(\phi)$. In particular, it was that $P_0(n)\sim \exp[-\theta_d n]$ for large $n$ with 
$\theta_d=\log [\pi/2]$. The persistence $P_0(n)$ was shown to 
be identical to the average fraction of metastable configurations (originally
computed in Ref. \cite{DG}) in the corresponding 
Ising spin glass chain\cite{MD}. 

The purpose of this paper is to show that $P_m(n)$ for any $m\ge 0$ can also be 
calculated exactly for the sequence in Eq. (\ref{psi}) and turns out to be universal.
Let us summarize our main results which are all independent of the distrbition $\rho(\phi)$: 

$\bullet$  We show that the mean number of 
sign changes upto $n$ steps (i.e., when the sequence size is $n+1$) is given by the exact formula, 
$\langle m\rangle =n/3$ for all
$n>0$.

$\bullet$ The variance is given by the formula, $\sigma^2_n=\langle m^2\rangle-{\langle 
m\rangle}^2=[16n+3+\delta_{n,1}]/90$
for all $n>0$. 

$\bullet$ We show that analogus to its continuous counterpart, the generating function 
${\tilde P}(p,n)=\sum_{m=0}^n 
p^m P_m(n)\sim \exp[-\theta_d(p)\,n]$ where the `discrete partial survival' exponent $\theta_d(p)$ is 
given by
the closed form expression 
\begin{equation}
\theta_d(p)= \log\left[{ { {\sin}^{-1}\left(\sqrt{1-p^2}\right)}\over {\sqrt{1-p^2}} }\right],\,\,\, 
0\le p\le 1.
\label{thetadp}
\end{equation}
This result can be analytically continued to $p\ge 1$.

$\bullet$ We also show that in the limit $m\to \infty$, $n\to \infty$ but keeping $x=m/n$ fixed,
$P_m(n)\sim \exp[-n\Phi(x)]$ where $\Phi(x)$ is a universal large deviation function
that we compute.

We start by defining $P^{\pm}_{m,n}(\phi_0)$ to be the joint probability that the first member 
of the sequence in Eq. (\ref{psi}) $\pm \psi_1>0$ and that the sequence undergoes $m$
sign changes upto $n$ steps, given the value of $\phi_0$. It is then easy to see that
they satisfy the following recursion relations
\begin{eqnarray}
P^{+}_{m,n+1}(\phi_0)&=&\int_{-\phi_0}^{\infty}d\phi_1 
\rho(\phi_1)F^{+}_{m,n}(\phi_1) \nonumber \\
P^{-}_{m,n+1}(\phi_0)&=&\int_{-\infty}^{-\phi_0}d\phi_1
\rho(\phi_1)F^{-}_{m,n}(\phi_1),
\label{recur1}
\end{eqnarray}
where $F^{\pm}_{m,n}(\phi_1)=P^{\pm}_{m,n}(\phi_1)+P^{\mp}_{m-1,n}(\phi_1)$
and the 
initial conditions are $P^{\pm}_{m,0}(\phi_0)=0$ for $m>0$,
$P^{+}_{0,0}(\phi_0)=\int_{-\phi_0}^{\infty}\rho(\phi_1)d\phi_1$
and $P^{-}_{0,0}(\phi_0)=1-P^{+}_{0,0}(\phi_0)$. The generating functions
${\tilde P}^{\pm}_n(p,\phi_0)=\sum_0^{\infty}P^{\pm}_{m,n}(\phi_0)p^m$ then satisfy
the recursion relations
\begin{eqnarray}
{\tilde P}^{+}_{n+1}(p,\phi_0)&=&\int_{-\phi_0}^{\infty}d\phi_1
\rho(\phi_1){\tilde F}^{+}_n(p,\phi_1) \nonumber \\
{\tilde P}^{-}_{n+1}(p,\phi_0)&=&\int_{-\infty}^{-\phi_0}d\phi_1
\rho(\phi_1){\tilde F}^{-}_n(p,\phi_1),
\label{recur2}
\end{eqnarray}                         
where ${\tilde F}^{\pm}_n(p,\phi_1)={\tilde P}^{\pm}_n(p,\phi_1)+p{\tilde P}^{\mp}_n(p,\phi_1)$ 
with the initial conditions, ${\tilde P}^{+}(p,\phi_0)=\int_0^{\infty}\rho(\phi_1)d\phi_1$
and ${\tilde P}^{-}(p,\phi_0)=1-{\tilde P}^{+}(p,\phi_0)$. Further simplification can be made
by differenting Eq. (\ref{recur2}) with respect to $\phi_0$ followed by a change of 
variable
$\phi_0\to u=\int_0^{\phi_0} \rho(\phi)d\phi$ and then using the symmetry 
$\rho(\phi)=\rho(-\phi)$.
Writing ${\tilde P}^{\pm}_n(p,\phi_0)={\tilde p}^{\pm}_n(u)$ (suppressing the $p$ dependence 
for 
convenience) we find two coupled non-local recursion relations
\begin{equation}
{ {d {\tilde p}^{\pm}_{n+1}(u)}\over {du}}=\pm\left[{\tilde p}^{\pm}_n(-u)+p{\tilde 
p}^{\mp}_n(-u)\right],
\label{recur3}
\end{equation}
with the initial conditions ${\tilde p}^{\pm}_0(u)=1/2\pm u$ and the boundary conditions
${\tilde p}^{\pm}(\mp 1/2)=0$. Note that the explicit dependence on the distribution
$\rho(\phi)$ disappears in Eq. (\ref{recur3}). As a result, all further quantities
computed from these recursion relations will be independent of $\rho(\phi)$ provided
$\rho(\phi)$ is symmetric and continuous.

In principle, one can solve the recursion relations in Eq. (\ref{recur2}) by the generating 
function method. However to calculate the mean and the variance of zero crossings, it is simpler to 
directly analyze Eq. (\ref{recur3}). Let $E^{\pm}_n(\phi_0)=\sum_{m=0}^{\infty}P^{\pm}_{m,n}(\phi_0)$ be
the expected number of sign changes upto $n$ steps with the first member $\psi_1$ positive (or 
negative) and given $\phi_0$. Let us write $E^{\pm}_n(\phi_0)=e^{\pm}_n(u)$ after
making the change of variable $\phi_0\to u$. The average number 
of crossings is then given by
$\langle m\rangle = \int_{-\infty}^{\infty}[E^{+}_n(\phi_0)+E^{-}_n(\phi_0)]\rho(\phi_0)d\phi_0
=\int_{-1/2}^{1/2}[e^{+}_n(u)+e^{-}_n(u)]du$.
Differentiating Eqs. (\ref{recur3}) once with respect to $p$ and putting $p=1$, we get
\begin{equation}
{ {de^{\pm}_{n+1}}\over {du}}=\pm\left[e^{+}_n(-u)+e^{-}_n(-u)\right]+u\pm {1\over {2}},
\label{recurm}
\end{equation}
with the initial conditions $e^{\pm}_0(u)=0$ and the boundary conditions $e^{\pm}_n(\mp 1/2)=0$.
These recursion relations can be solved exactly and one gets
$e^{\pm}_{n+1}=(1/2\pm u)(n-4)/12\pm u^3/3+u^2/2\pm u/2+1/6$. Hence  
$e_n(u)=e^{+}_n(u)+e^{-}_n(u)=u^2+n/3-1/12$. This then gives the exact result $\langle m\rangle 
=\int_{-1/2}^{1/2}e_n(u)du=n/3$ for all $n\ge 0$, independent of 
$\rho(\phi)$.

The variance $\sigma^2_n=\langle m^2\rangle -{\langle m\rangle}^2$ can be computed in a similar way
by differentiating Eq. (\ref{recur3}) twice with respect to $p$, putting $p=1$ and solving the
resulting recursion relations. The functions
$G^{\pm}_n(\phi_0)=\sum_{m=0}^{\infty}m(m-1)P^{\pm}_{m,n}(\phi_0)=g^{\pm}_n(u)$ satisfy
the following inhomogeneous recursion relations for $n>0$,
\begin{equation}
{ {dg^{\pm}_{n+1}}\over {du}}=\pm \left[g^{+}_n(-u)+g^{-}_n(-u)\right]\pm 2e^{\mp}_n(-u),
\label{recurv}
\end{equation}
with the initial conditions $g^{\pm}_0(u)=g^{\pm}_1(u)=0$ and boundary conditions $g^{\pm}_n(\mp 
1/2)=0$. Using the known values of $e^{\pm}_n(u)$ one can again solve Eq. (\ref{recurv}) 
explicitly. This finally gives $\langle m(m-1)\rangle=\int_{-1/2}^{1/2}[g^{+}_n(u)+g^{-}_n(u)]du=
(10n^2-14n+3)/90$ for all $n\ge 2$. Using $\langle m\rangle =n/3$, one gets
$\sigma^2_n=[16n+3+\delta_{n,1}]/90$ for all $n>0$, again independent of 
$\rho(\phi)$.

We now turn to the calculation of the partial survival exponent. We expect that for large $n$,
${\tilde p}^{\pm}_n(u)\approx {\lambda}^{-n}f^{\pm}(u)$ where $\lambda=\exp[\theta_d(p)]$.
Substituting this asymptotic form in Eq. (\ref{recur3}) we get the non-local eigenvalue equation
\begin{equation}
{ {df^{\pm}(u)}\over {du}}=\pm \lambda\left[f^{+}(-u)+f^{-}(-u)\right], 
\label{eigen}
\end{equation}
subject to the two boundary conditions, $f^{+}(- 1/2)=0$ and $f^{-}(1/2)=0$.  
Diagonalizing Eq. (\ref{eigen}) and solving the resulting non-local equations
we get the most general solutions of Eq. (\ref{eigen})
\begin{equation}
f^{\pm}(u)= A^{\pm}\cos(\mu u)+ B^{\pm} \sin(\mu u), 
\label{fu}
\end{equation}
where $\mu=\lambda\sqrt{1-p^2}$ and the four constants $A^{\pm}$ and $B^{\pm}$
can be written in terms of only two unknown constants $a$ and $b$ via the relations,
$A^{+}=ap$, $A^{-}=b\sqrt{1-p^2}-a$, $B^{+}=bp$ and $B^{-}=a\sqrt{1-p^2}-b$.
The solution in Eq. (\ref{fu}) must satisfy the two boundary conditions
$f^{\pm}(\mp 1/2)=0$ which gives two homogeneous linear equations for the
unknown constants $a$ and $b$. Eliminating $a$ and $b$ from these two equations
one gets $\mu={\sin}^{-1}[\sqrt{1-p^2}]$ and hence
$\lambda={\sin}^{-1}[\sqrt{1-p^2}]/\sqrt{1-p^2}$. Using the relation 
$\theta_d(p)=\log \lambda$, we obtain the result in Eq. (\ref{thetadp}), once
again independent of $\rho(\phi)$. Note that for $p= 0$, $\theta_d(p)$
reduces to the usual discrete persistence exponent $\theta_d=\log (\pi/2)$.

The expression for $\theta_d(p)$ in Eq. (\ref{thetadp}) is valid in the range $0\le p \le 1$. 
However, in principle, one 
can define the generating function ${\tilde P}(p,n)=\sum_{m=0}^{\infty}P_m(n)p^n$ even 
for $p>1$. Then for $p>1$ one expects ${\tilde P}(p,n)$ to diverge as $n\to \infty$
indicating $\theta_d(p)$ becomes negative for $p>1$. Indeed one can easily get the result for $p>1$
by analytically continuing the expression in Eq. (\ref{thetadp}) to the range $p\ge 1$
and this gives
\begin{equation}
\theta_d(p)=\log\left[ { { \log\left(p+\sqrt{p^2-1}\right)}\over {\sqrt{p^2-1}} }\right].
\label{thetadpg}
\end{equation}
Thus $\theta_d(p)$ tends to $-\infty$ rather slowly as $\theta_d(p)\sim \log[\log(2p)/p]$
as $p\to \infty$.

We next analyse the distribution $P_m(n)$ in an interesting scaling limit.
Since the average number of crossings scale linearly with the size $n$ as $\langle 
m\rangle =n/3$, a natural scaling limit is when $m\to \infty$, $n\to \infty$ but keeping
the ratio $x=m/n$ fixed. In this limit, we show that $P_m(n)\sim \exp[-n \Phi(x)]$ where
$\Phi(x)$ is a large deviation function which is universal, i.e., independent of the
distribution $\rho(\phi)$. Large deviation functions associated with different
physical observables have appeared before in
the context of various nonequilibrium systems\cite{DL}. The present model
provides an example where there is a large deviation function associated
with the number of zero crossings that can be 
computed explicitly. Indeed, substituting the ansatz $P_m(n)\sim \exp[-n \Phi(x)]$
in the generating function one gets ${\tilde P}(p,n)\sim \sum_{m=0}^{\infty}
P_m(n)p^m \sim \int_0^{\infty} dx \exp[-n\{\Phi(x)-x\log p\}]$. 
In the large $n$
limit the integral can be evaluated by the steepest descent method and one gets            
${\tilde P}(p,n)\sim \exp[-n G(p)]$ where $G(p)={\rm min}_x\{\Phi(x)-x\log p\}$. 
On the other hand, by definition, ${\tilde P}(p,n)\sim \exp[-\theta_d(p)n]$ for large $n$.
This establishes the relation, ${\rm min}_x\{\Phi(x)-x\log p\}=\theta_d(p)$.
Thus $\theta_d(p)$ is just the Legendre transform of $\Phi(x)$.
Inverting this Legendre transform we get
\begin{equation}
\Phi(x)= {\rm max}_p\left[x \log p +\theta_d(p)\right],
\label{phix}
\end{equation}
where $\theta_d(p)$ is given exactly by Eqs. (\ref{thetadp}) and (\ref{thetadpg})
in the range $p\ge 0$. Note that determining $\Phi(x)$ from Eq. (\ref{phix}) requires a knowledge
of $\theta_d(p)$ not just in the range $0\le p\le 1$ but also for $p\ge 1$.
Thus we will need both the formulas in Eqs. (\ref{thetadp}) and (\ref{thetadpg}).
\begin{figure}
\narrowtext\centerline{\epsfxsize\columnwidth \epsfbox{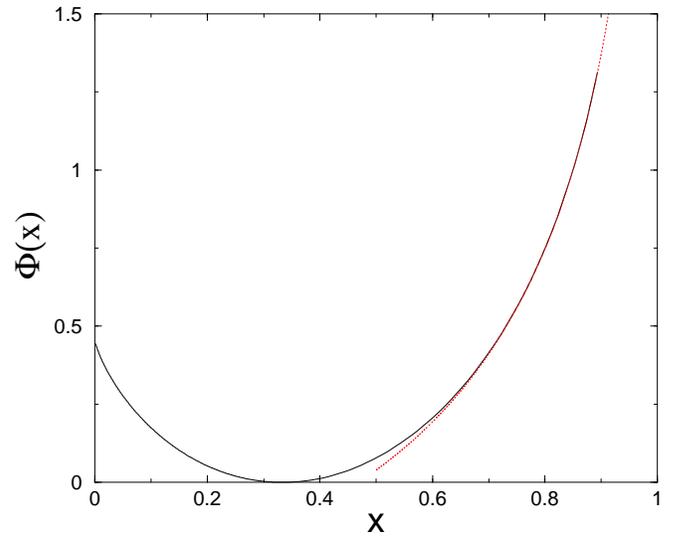}}
\caption{The large deviation function $\Phi(x)$ plotted against $x$. The solid line represents 
the function obtained using Mathematica. The dotted line represents the
analytical asymptotic form $\Phi(x)=-\log(1-x)-1+(1-x)\log 2$ in the limit $x\to 1$.
The function $\Phi(x)\to \log (\pi/2)=0.451583\ldots$ as $x\to 0$.}
\end{figure}

We have obtained $\Phi(x)$ from Eq. (\ref{phix}) using Mathematica and is displayed in Fig. 1.
Since the number of crossings $m\le n$, the allowed range of $x$ is $0\le x\le 1$.
One can 
analytically determine the behavior of $\Phi(x)$ in the three limits $x\to 0$, $x\to 1$
and $x\to 1/3$. First consider the limit $x\to 0$. This corresponds to $p\to 0$ limit
of $\theta_d(p)$. Expanding Eq. (\ref{thetadp}) for small $p$, we get $\theta_d(p)\approx
\log(\pi/2-p)$. Substituting this in Eq. (\ref{phix}) and maximizing with respect to $p$
gives $\Phi(x)\approx \log(\pi/2) + x\log (x)$ as $x\to 0$. Next consider the opposite limit when
$x\to 1$. This limit correspond to $\theta_d(p)$ in the limit $p\to \infty$. Hence
we need to now use the analytically continued formula in Eq. (\ref{thetadpg}). Expanding
Eq. (\ref{thetadpg}) for large $p$, we get $\theta_d(p)\approx \log[\log(2p)/p]$ to leading
order. Substituting this asymptotic form in Eq. (\ref{phix}) and maximizing with respect
to $p$ we get $\Phi(x)\approx -\log(1-x)-1+(1-x)\log 2$ as $x\to 1$. In Fig. 1, this
asymptotic form is shown by the dotted line to which $\Phi(x)$ approaches rather quickly
as $x\to 1$.

The most interesting limit, however, is when $x\to 1/3$, i.e. $m\to \langle m\rangle$.
This limit in $x$ corresponds to $p\to 1$ limit of $\theta_d(p)$. It is easy to see that
both the limits $p\to 1^{-}$ and $p\to 1^{+}$ yield the same result. Let us consider
the case when $p=1-\epsilon$ where $\epsilon\to 0$. Expanding Eq. (\ref{thetadp}) 
in powers of $\epsilon$, we get $\theta_d(p)\approx
\epsilon/3 + 7\epsilon^2/90 + O(\epsilon^3)$. Substituting this in Eq. (\ref{phix})
and maximizing with respect to $p=1-\epsilon$, we get as $x\to 1/3$,
\begin{equation}
\Phi(x)\approx {{45}\over {16}}{\left( x-{1\over {3}}\right)}^2.
\label{clt}
\end{equation}
This limiting form can also be derived independently from a central limit theorem.
To see this we write the number of sign changes $m$ as the sum 
$m=\sum_{i=1}^n w_i$ with $w_i=1-\theta(\psi_i\psi_{i+1})$ and
$\theta(x)$ is the Heaviside step function. Thus $m-\langle m\rangle=\sum_{i=1}^n (w_i-\langle 
w_i\rangle)$. Clearly in the limit $m\to \langle m\rangle$, the variables $(w_i-\langle 
w_i\rangle)$ become only weakly correlated. Then in the limit when $n$ is much larger than the 
correlation length between these variables one expects the central limit theorem to hold
predicting a Gaussian distribution for $m$, $P_m(n)\sim \exp[-(m-\langle 
m\rangle)^2/{2\sigma^2_n}]$.
Using the already derived results $\langle m\rangle=n/3$ and $\sigma^2_n\approx 8n/45$ for large 
$n$,
we find $P_m(n)\sim \exp[-45n(x-1/3)^2/16]$ thus yielding the same
$\Phi(x)$ as in Eq. (\ref{clt}). Thus this limit provides an indipendent check
of our results for the mean and the variance. The three limiting behaviors of $\Phi(x)$ are 
summmarized
as follows
\begin{equation}
\Phi(x)\approx\cases{
\log(\pi/2)+x\log x        &$x\to 0$,\cr
{{45}\over {16}}{(x-{1\over {3}})}^2    &$x\to 1/3$,\cr
-\log(1-x)-1+(1-x)\log 2   &$x\to 1$. \cr}
\label{limiting}
\end{equation} 

We conclude with a discussion of the implications of our results for an Ising spin glass
chain described by the Hamiltonian, $H=-\sum J_{i,i+1}s_is_{i+1}$ with $s_i=\pm 1$ and
the bonds $J_{i,i+1}$'s are i.i.d random variables each drawn from the same symmetric and
continuous distribution. A spin will be called metastable if the cost of energy to flip
it under zero temperature Glauber dynamics is positive, i.e. $\Delta 
E_i=2s_i[J_{i-1,i}s_{i-1}+J_{i,i+1}s_{i+1}]>0$. A given spin configuration (with fixed
$J$'s) consists of alternate domains of metastable and non-metastable spins. A natural
question is what is the average (over disorder) probability $P(m,n)$ that there are $m$ such 
domains in a chain
of length $n$. Defining the new variables $\phi_i=2J_{i,i+1}s_is_{i+1}$ which
are also i.i.d variables, we see that the energy costs $\Delta E_i=\phi_{i}+\phi_{i-1}$
form exactly the sequence studied in this paper. The average domain number 
probability $P(m,n)$
is then identical to the probability of having $2m$ sign changes in the sequence
$\{\Delta E_i\}$ upto $n$ steps, i.e., $P_{2m}(n)$ that has been computed exactly in this 
paper. Clearly for $m=0$, $P(0,n)=P_0(n)$ is just the fraction of fully metastable
configurations (out of the $2^n$ configurations) at zero temperature and is the same
as the persistence of the sequence $\{\Delta E_i\}$\cite{MD}. Note that one
can easily generalize this question of the domain number probability to higher
dimensions as well. The study of the statistics of the domains of metsatable spins 
in higher dimensional spin glasses may provide interesting insights into the nature
of the low temperature phase.

I thank D. Dhar and A.J. Bray for useful discussions.

\end{multicols}

\end{document}